\documentclass[aps,twocolumn,showpacs,superscriptaddress,amsmath,amssymb,amsfonts,floatfix,longbibliography]{revtex4-1}

\usepackage{graphicx}
\usepackage{dcolumn}
\usepackage{bm}
\usepackage[dvipsnames]{xcolor}

\usepackage{microtype}
\usepackage{subfigure}
\usepackage{wrapfig}
\usepackage{cancel}
\usepackage{color}
\usepackage{textcomp}
\usepackage{amsmath}
\usepackage{amsfonts}
\usepackage{amssymb}
\usepackage{array}

\newcommand{\rev}[1]{\textcolor{black}{#1}} 
\newcommand{\CS}{CaSb$_2$}

\newcommand{\ChiV}{$\chi_{\text{V}}$}

\newcommand{\Tc}{$T_{\text c}$}

\newcommand{\mHct}{$\mu_{\rm{0}} H_{\rm{c2}}(0)$} 
\newcommand{\mHctwo}{$\mu_{\rm{0}} H_{\rm{c2}}$} 
\newcommand{\mHco}{$\mu_{\rm{0}} H_{\rm{c1}}(0)$}
\newcommand{\mHsco}{$\mu_{\rm{0}} H^{\rm{*}}_{\rm{c1}}(0)$}
\newcommand{\mHc}{$\mu_{\rm{0}} H_{\rm{c}}(0)$}
\newcommand{\mSR}{$\mu\rm{SR}$}

\begin{document}

\preprint{APS/123-QED}

\title{Time-Reversal Symmetry Breaking Superconductivity in \CS\ }


\author{M. Oudah$^{1,2}$}
 \email{mohamed.oudah@ubc.ca}
\author{Y. Cai$^{1,2}$}%
\author{M. V. De Toro Sanchez$^{1}$}%
\author{J. Bannies$^{1,2,3}$}%
\author{M. C. Aronson$^{1,2}$}%
\author{K. M. Kojima$^{1,2}$}%
\author{D. A. Bonn$^{1,2}$}%
\affiliation{%
$^1$Stewart Blusson Quantum Matter Institute, University of British Columbia, Vancouver, British Columbia V6T 1Z4, Canada\\
$^2$Department of Physics and Astronomy, University of British Columbia, Vancouver, Canada V6T 1Z1, Canada\\
$^3$Department of Chemistry, University of British Columbia, Vancouver, Canada V6T 1Z1, Canada\\
}%


\date{\today}

\begin{abstract}
\rev{\CS\ is a bulk superconductor and a topological semimetal, making it a great platform for realizing topological superconductivity. In this work, we investigate the superconducting upper and lower critical field anisotropy using magnetic susceptibility, and study the superconducting state using muon spin-relaxation. The temperature dependence of transverse-field relaxation rate can be fitted with a single-gap model or two-gap model.  Zero-field relaxation shows little temperature dependence when the muon-spin is parallel to the $c*$-axis, while an increase in relaxation appears below 1 K when the muon-spin is parallel to the $ab$-plane. We conclude an $s+is$ order parameter considering the breaking of time-reversal symmetry (TRS), which originates from competing interband interactions between the three bands of \CS . To explain the direction-dependent breaking of TRS we suggest loop currents developing in the plane of distorted square-net of Sb atoms.}



\end{abstract}

\pacs{Valid PACS appear here}

\maketitle

When a material enters the superconducting state it breaks U(1) gauge symmetry, and breaking of any additional symmetries is typically an indication of unconventional superconductivity~\cite{sigrist1991phenomenological}.
In some unconventional superconductors, time-reversal symmetry (TRS) is broken as the material enters the superconducting state as proven by detection of spontaneous magnetic fields below the onset of superconductivity. This spontaneous magnetic field has been detected in zero-field muon relaxation measurements in unconventional superconductors such as UPt$_3$~\cite{luke1993muon} and Sr$_2$RuO$_4$~\cite{luke1998time,xia2006high}.
These materials are expected to host chiral ground states with broken mirror and time-reversal symmetries, a topological superconducting state~\cite{yanase2017mobius,kallin2012chiral}.
\rev{When present, the breaking of TRS in the superconducting state typically appears in materials containing magnetic elements or strong spin-orbit coupling (SOC). \CS\ presents an alternative direction for unconventional superconductivity; it does not contain any magnetic elements nor does it have strong SOC, but is a Dirac semimetal with topologically non-trivial electronic states protected by non-symmorphic symmetry of the crystal structure.}

Spontaneous magnetic fields can emerge in non-centrosymmetric superconductors, where the lack of inversion symmetry results in a spin-split Fermi surface due to antisymmetric SOC and a singlet-triplet mixing in the superconducting state~\cite{gor2001superconducting}, such as LaNiC$_2$~\cite{hillier2009evidence}, Re$_6$Zr~\cite{singh2014detection}, and La$_7$Ir$_3$~\cite{barker2015unconventional}.
This breaking of TRS can even appear in centrosymmetric multi-gap superconductors, for specific compositions, such as the case in FeSe~\cite{watashige2015evidence}, Fe(Se,Te)~\cite{farhang2023revealing} and Ba$_{1-x}$K$_x$Fe$_2$As$_2$~\cite{grinenko2017superconductivity,grinenko2020superconductivity} and single $s$-wave gap locally non-centroscymmetric SrPtAs and CaPtAs with strong SOC~\cite{biswas2013evidence, shang2020simultaneous}. In PrOs$_4$Sb$_{12}$ TRS breaking appears below \Tc\ and is discussed in relation to nonmagnetic quadrupolar fluctuations in the normal state~\cite{aoki2003time}.
In the Sr-doped topological insulator Bi$_2$Se$_3$, TRS breaking has been discussed in relation to the anisotropic Dirac cone dispersion in the normal state band structure of doped Bi$_2$Se$_3$ allowing for triplet pairing~\cite{neha2019time}. In all of the above materials, the onset of TRS breaking coincides with \Tc\ under ambient conditions. In \CS\ we will show a new TRS phenomenon, appearing below \Tc\ and only detected when muon spins are in a specific direction.


\begin{figure*}[t]
\centering
\includegraphics[width=170mm]{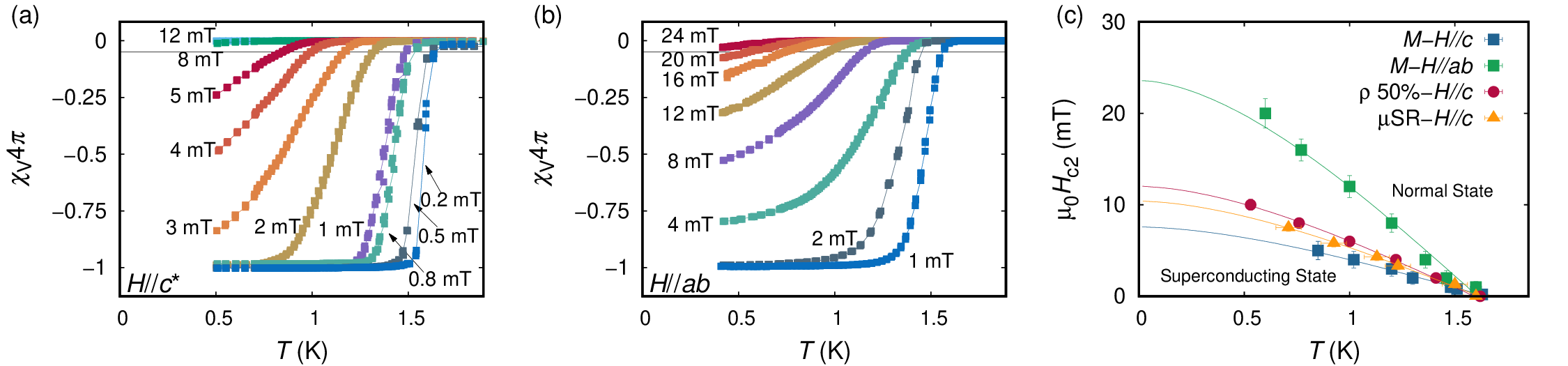}
\caption{Temperature dependence of the dc susceptibility measured with different applied fields using a zero-field-cooling procedure measured with $H//c*$ and $H//ab$ shown in (a) and (b), respectively. Data represented as volume susceptibility, \ChiV , and data in (b) are from Ref~\cite{oudah2022superconductivity}. (c) The temperature dependence of the upper critical field \mHctwo\ determined from \ChiV\ with $H//c*$ and $H//ab$, from resistivity~\cite{oudah2022superconductivity}, and from TF-\mSR\ with $H//c*$. Solid lines are WHH fits described in the text to the data used to estimate \mHct .}
\label{Aniso}
\end{figure*}

\CS\ belongs to the family of non-symmorphic antimonides $M$Sb$_2$ ($M$= Alkaline-Earth, Rare-Earth) containing screw rotation symmetry. \CS\ is a compensated semimetal~\cite{funada2019spin,oudah2022superconductivity}. Its calculated Fermi-surface supports \CS\ being a topological nodal-line semimetal~\cite{oudah2022superconductivity,chuang2022fermiology} due to the non-symmorphic space group 11, $P2_1/m$~\cite{deller1976darstellung}. The compensated semimetal state is related to three bands crossing the Fermi level, two electron-like bands dominated by contributions from the Sb site forming a distorted square-net and a hole-like band dominated by contributions from the other Sb site forming a zig-zag chain along the $b$-direction~\cite{oudah2022superconductivity}. Superconductivity was discovered recently in polycrystalline samples~\cite{ikeda2020superconductivity}, and further confirmed in single crystal samples~\cite{oudah2022superconductivity}. Recently, the anisotropy of the upper critical field of \CS\ based on resistivity measurements and the lower critical field estimate based on magnetization at 0.55~K have been reported~\cite{ikeda2022quasi}. The specific heat transition in single crystal samples suggests deviation from a single $s$-wave gap, with the possibility of multiple gaps in the superconducting state~\cite{oudah2022superconductivity}. A coherence peak was observed near \Tc\ in Sb-nuclear quadrupole resonance (NQR) suggesting $s$-wave superconductivity~\cite{takahashi2021s}, at least near \Tc . Using a tunnel diode oscillator (TDO), the temperature dependence of the penetration depth in \CS\ reveals the presence of multiple gaps and exponential behaviour at low temperature that indicates nodeless superconductivity~\cite{duan2022nodeless}. These reports warrant further investigations of the anisotropy of the superconducting state of \CS\ and studying the superconductivity using a local probe such as muon spin rotation/relaxation (\mSR ).

Here we report the critical field anisotropy from magnetic susceptibility, estimate the penetration depth and the temperature dependence of the superconducting gap from transverse-field (TF) muon data, and find evidence for TRS breaking below \Tc\ in zero-field (ZF) \mSR\, but only when muon spin lies in the $ab$-plane, showing that a spontaneous field perpendicular to the $ab$-plane emerges in \CS . Considering that three bands cross the Fermi level of \CS , the TRS breaking in ZF-\mSR\ and TF-\mSR\ can be reconciled by considering a complex order parameter such as $s+is$. The breaking of TRS is detected below $\sim1$~K, which is well below the \Tc\ of 1.6~K.
The observation of TRS breaking despite the lack of magnetic elements or strong spin-orbit-coupling in \CS\ is new, and may be related to the topological nature of the superconducting state arising from the three bands with non-trivial topology crossing the Fermi level.


\section{Results and Discussion}

\begin{figure*}[t]
\includegraphics[width=178mm]{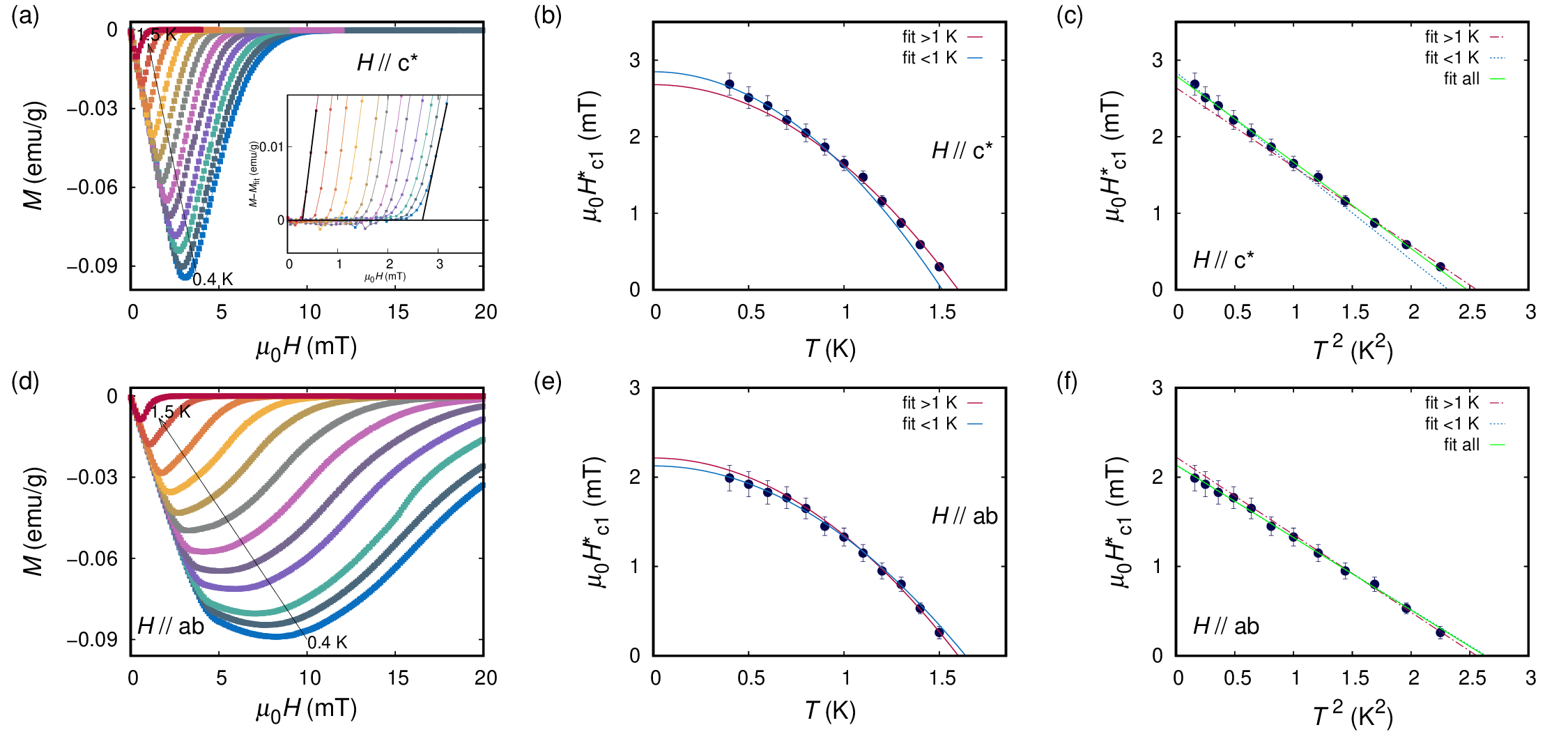}
\caption{The magnetization ($M$) as a function of applied field $\mu_0H$ measured at different temperatures $T$ below \Tc\ with field applied parallel to $c*$ direction and $ab$-plane shown in (a) and (d), respectively. \rev{Measurements were performed on the same sample, but changing its orientation with respect to the applied field.} A degaussing procedure was carried out between measurements, and a linear fit was applied to the low field region of the 0.4~K data. Lower critical field \mHsco\ as a function of temperature ($T$) for $H//c*$ and $H//ab$, shown in (b) and (e), estimated using the magnetization data in (a) and (d) by subtracting the linear fit to 0.4~K data from all the curves, shown in inset of (a). Linear fit is applied to the upturn data and the intersect is defined as \mHsco .  (c) and (f) show \mHsco\ as a function of $T^2$ and fits to the data $<1$~K (dashed blue line), $>1$~K (dashed red line), and over the entire temperature range are shown (solid green line).}
\label{Hc1}
\end{figure*}

\subsection{\label{Anisotropy}Critical Field Anisotropy}

Measurements of the anisotropy of the field-dependent dc magnetic susceptibility around the superconducting transition were performed down to 0.4~K, as shown in Fig.~\ref{Aniso}(a) and (b). The temperature dependent magnetic susceptibility was measured for fields applied parallel or perpendicular to the plate of the crystal, where the $ab$-plane lies in the plane of plate-like crystals. We define $c*$ as the direction perpendicular to the $ab$-plane for this monoclinic crystal structure. In both cases, the transition temperature was defined as the 5\% volume drop in the volume susceptibility where the 100\% volume was defined as the signal in the lowest field measured (1~mT). The demagnetization correction based on the Brandt formula~\cite{brandt1999irreversible} yields a 100\% volume fraction for the case of $H//ab$, but the same correction results in a 30\% volume fraction in the case of $H//c*$. This is due to our plate-like crystals not being perfect rectangular slabs, as assumed in the calculation of the demagnetization factor. For the $H//c*$ direction, we take the 5\% volume drop relative the signal measured in the lowest field being the 100\% volume fraction, as shown in Fig.~\ref{Aniso}(a). \rev{The volume susceptibility without demagnetization correction is shown in Fig.~\ref{UncorrectedVolumeSusc}.}

The upper critical field was estimated using the Werthamer-Helfand-Hohenberg (WHH) relation~\cite{werthamer1966temperature} for the data as shown with the lines in Fig.~\ref{Aniso}(c). The temperature dependence of $H_{c 2}^{a b}$ is typical of a type-II superconductor, as the transition moves to lower and lower field values upon increasing the temperature. Similar behavior is observed for measurements with the other field orientation $H_{c 2}^{c}$. The estimated upper critical fields are $H_{c 2}^{c}=8.1 \mathrm{~mT}$ and $H_{c 2}^{ab}=24.7 \mathrm{~mT}$, thus yielding an anisotropy ratio $\gamma_{\text {anisotropy}}=H_{c 2}^{a b} / H_{c 2}^{c}$ in \CS\ of about 3.1. We estimate the coherence length based on the upper critical field, and obtain 202~nm and 66~nm for $\xi_{\rm{GL},ab}$ and $\xi_{\rm{GL},c}$, respectively.
We note that deviation from WHH relation near \Tc , where \mHct\ increases slowly with decreasing temperature, is consistent with $s$-wave superconductivity in a nodal-line semimetal~\cite{endo2023theoretical}. The temperature dependence of the anisotropy of the upper critical field for \CS , and a fit based on model for $s$-wave superconductivity in an anisotropic nodal-line semimetal are shown in the Sup.~Fig.~\ref{Hc2_nodal-line}.


To estimate the lower critical field \mHco\ we measured the field-dependent magnetization ($M$) at different temperatures below the critical temperature in both directions, as shown in Fig.~\ref{Hc1}(a) and (d).
We use a linear fit to the low-field data measured at 0.4~K in each direction, and subtract this fit from all the measured curves for measurements in each direction, as demonstrated in the inset of Fig.~\ref{Hc1}(a). The value of the uncorrected \mHsco\ at each temperature was determined based on the intersect of a linear fit to the upturn of this subtracted data, $M-M_{fit}$, and the horizontal line where $M-M_{fit}=0$ marked in the inset of Fig.~\ref{Hc1}(a) for 0.4~K and 1.5~K. Previously, we estimated \mHsco\ from the subtracted data by setting a threshold for deviation from zero~\cite{oudah2022superconductivity}, but we find this method overestimates \mHsco\ in the $H//c*$ direction, as shown for a threshold of $M-M_{fit}=0.001~\rm{emu/g}$ in Sup.~Fig.~\ref{Hc1_Multi}. The \mHsco\ plotted against the corresponding temperature for the $H//c*$ and $H//ab$ direction are shown in Fig.~\ref{Hc1}(b) and (e), and the same data plotted against temperature squared are shown in Fig.~\ref{Hc1}(c) and (f). We note that the temperature dependence of \mHsco\ is fitted well by the equation~\cite{powell1978definitions}
\begin{equation}
    \mu_0 H^*_{c1}(T) = \mu_0 H^*_{c1}(0) \left[1- \left(\frac{T}{T_c}\right)^2 \right]\
    \label{MuHcl}
\end{equation}

\noindent
where \mHsco\ is the lower critical field at 0~K, where fits are shown with green line in Fig.~\ref{Hc1}(c) and (f). Although the data fit the equation above well, we attempt two independent fits to the temperature regions above and below 1~K in search of any anomaly that can be reconciled with our zero-field \mSR . 
Enhancement of the lower critical field was observed in UPt$_3$~\cite{amann1998magnetic} and PrOs$_4$Sb$_{12}$~\cite{cichorek2005pronounced}, and this was related to the emergence of a second superconducting phase at low temperature. We discuss this further below in Sec.~\ref{ZF}.


\begin{figure*}[t]
\includegraphics[width=178mm]{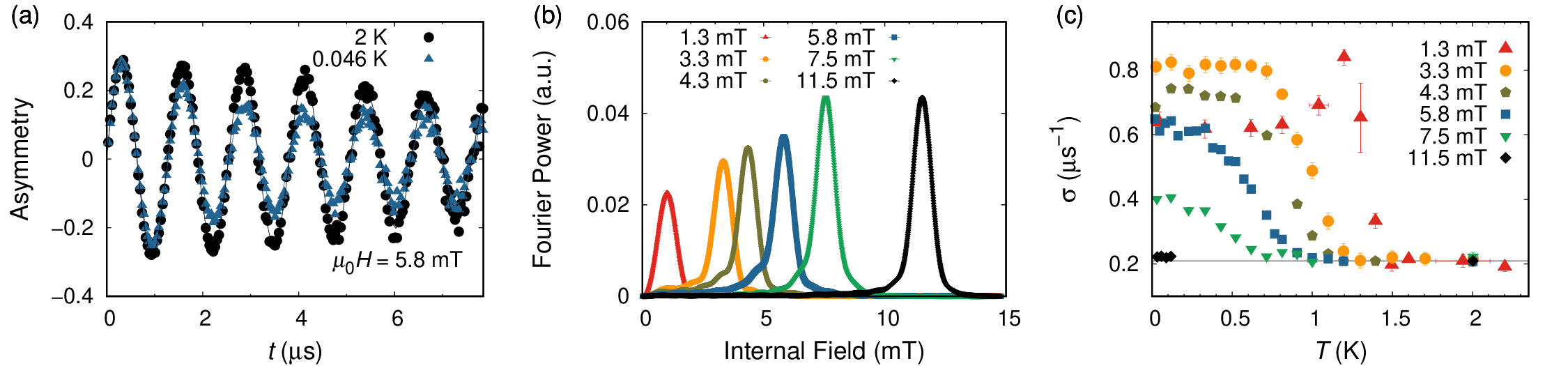}
\caption{Representative TF-\mSR\ signals collected above and below \Tc\ in \CS\ under an applied magnetic field of 5.8~mT. The solid lines are fits using the sinusoidal decaying function described in the text. (b) The Fourier transform of the M15 \mSR\ Asymmetry for measurements in different applied magnetic fields at $\sim30$~mK, representing the field distribution of the local field probed by muons. The sharp peaks in the data indicate the applied field experienced by muons stopping in the silver cold finger. The broad features at lower/higher internal fields represents the muons stopping in the superconducting \CS\ sample. (c) The muon Gaussian relaxation rate $\sigma$ as a function of temperature in different applied magnetic fields.}
\label{TF-muon}
\end{figure*}

The typical equations used for estimating the penetration depth ($\lambda$) from \mHco\ do not have a solution for the values measured in our experiment. Instead, we estimate the \mHc\ from \mHco\ and \mHct\ with the following equation~\cite{tinkham2004introduction}:

\begin{equation}
    \mu_0H_c= \sqrt{\mu_0 H_{c1} \times \mu_0 H_{c2}}
    \label{MuHc}
\end{equation}

\noindent
Here we assume the \mHco\ values for $H//c*$ and $H//ab$ are equivalent to the \mHsco\ measured, without applying any demagnetization correction. This is due to the difficulty of calculating the demagnetization correction for the $H//c*$. 
The value of thermodynamic critical field \mHc\ is expected to be equivalent for both directions, but does not match for the current measurements on our samples. 
We take the average in both directions and estimate \mHc\ $=6.0\pm0.5$~mT, as shown in Table~\ref{SCAniso}, which is consistent with the value previously reported using the integral of $M(H)$ curve~\cite{ikeda2022quasi}.  Despite this consistency of \mHc , we suspect that the currently measured lower critical field values are inaccurate and should be the subject of future studies.
We estimate the Ginzburg-Landau parameter for each direction with the following equation:

\begin{equation}
    \mu_0 H_{c2} = \sqrt{2}\kappa_{GL}\mu_0H_c 
    \label{kappa}
\end{equation}

\noindent
where $\kappa_{GL}$ is the Ginzburg-Landau parameter.

We summarize our characterization of the superconducting state based on magnetic susceptibility measurements in Table~\ref{SCAniso}.
We estimate the upper critical field $H//c*$ extracted from the 50\% drop in resistivity measurement, previously published by some of the current authors~\cite{oudah2022superconductivity}, to be about $11.5\pm0.8$~mT in Fig.~\ref{Aniso}. This value is different from that reported recently by another group~\cite{ikeda2022quasi}, which may be due to different sample quality or due to different current densities in the resistivity measurements. Nevertheless, the anisotropy of the upper critical field of $\sim3$ from the resistivity measurements reported~\cite{ikeda2022quasi} is consistent with our anisotropy estimates based on magnetic susceptibility measurements. Here we note that for $H//c*$, it seems that the superconducting state of \CS\ is at the border of type-I/type-II regime. This poses challenges for our muon measurement as highlighted below.

\begin{table}[t]
\centering
\caption{Superconducting parameters derived from our measurements of \CS. \mHco\ is the average of that calculated for each direction based on Eq.~\ref{MuHc}.} 
\label{SCAniso}
\setlength{\extrarowheight}{8pt}
\begin{tabular}{|c c c|}
\hline
\textbf{Direction}   & \textit{ab} & \textit{c}* \\
\hline
\mHco    &  $2.1\pm0.4$~mT & $2.8\pm0.4$ \\
\mHct    & $24.7\pm0.8$~mT & $8.1\pm0.8$~mT \\
\mHc     & \multicolumn{2}{c|}{$6.0\pm0.5$~mT}  \\
$\xi_{\rm{GL}}$ & 202~nm & 66~nm \\
$\kappa_{\rm{GL}}$ & $2.90\pm0.26$ & $0.95\pm0.12$   \\
\hline
\end{tabular}
\end{table}

\subsection{\label{TF}Transeverse-Field Muon Study}

We perform transverse field \mSR\ measurements on \CS\ with applied magnetic field perpendicular to the $ab$-plane, which can be used to determine the temperature dependence of the penetration depth. With the temperature dependence of the penetration depth we can infer properties of the energy gap of the superconducting state of \CS . For these measurements we extract the field from the precession of muons inside the background silver, which is the most precise way of measuring the applied field. We performed the measurement with applied fields ranging from $1.3-11.5$~mT, and all measurements were performed on samples cooled in the applied field. We employed a beam of muons polarized such that the spin is perpendicular to their momentum, while the applied field is parallel to the momentum. The spin of the muon precesses with a frequency proportional to the local magnetic field, and, upon decay, emits a positron preferentially in the muon spin direction. Typically with a field that is above the lower critical field, we expect a well ordered flux line lattice when using a field cooling procedure. Typical time evolution of the asymmetry for \CS\ is shown in Fig.~\ref{TF-muon}(a), measured in 5.8~mT at 2.00~K and 0.03~K, above and below \Tc\ respectively. 

In the mixed state, we have an inhomogeneous field distribution due to the presence of a flux line lattice (FLL), which results in a decay of the precession signal as a function of time. We fit the asymmetry spectra using a two term sinusoidal decaying function

$$
\begin{aligned}
G_{\mathrm{TF}}(t) & =A\left[F \exp \left(\frac{-\sigma^{2} t^{2}}{2}\right) \cos \left(\omega_{1} t+\phi\right)\right. \\
& \left.+(1-F) \exp (-\psi t) \cos \left(\omega_{2} t+\phi\right)\right]
\end{aligned}
$$
\noindent
where the first term captures the signal from muons stopping in the sample and the second term captures the signal from muons stopping in the silver sample holder. $F$ is the fraction of the signal coming from the sample, while $\omega_{1}$ and $\omega_{2}$ are the muon precession frequencies in the sample and the background, respectively. The $A$ term is the total asymmetry and the $\phi$ is the initial phase of the muons. $\sigma$ and $\psi$ are the depolarization rates for the sample and the background signals, respectively. The $\sigma$ term contains a contribution from the field distribution caused by the vortex lattice in the superconducting state $\left(\sigma_{s c}\right)$ and the smaller, temperature independent, contribution from randomly oriented nuclear dipole moments $\left(\sigma_{N}\right)$. These two signals are added in quadrature, such that the contribution from the FLL can be obtained as $\sigma_{s c}=$ $\sqrt{\sigma^{2}-\sigma_{\mathrm{N}}^{2}}$. 
The superconducting relaxation rate $\left(\sigma_{s c}\right)$ indicates the mean square inhomogeniety in the field experienced by muons, $\left\langle\left(\Delta B)^{2}\right\rangle\right.$, due to the FLL~\cite{brandt1988magnetic}, where $\left\langle(\Delta B)^{2}\right\rangle=\left\langle(B-\langle B\rangle)^{2}\right\rangle$, which results in the relaxation rate for the FLL 

$$
\sigma_{s c}^{2}=\gamma_{\mu}^{2}\left\langle(\Delta B)^{2}\right\rangle
$$
\noindent
where $\gamma_{\mu}(=2 \pi \times 135.5 \mathrm{MHz} / \mathrm{T})$ is the muon gyromagnetic ratio. The Fourier power against internal magnetic field, shown in Fig.~\ref{TF-muon}(b), shows a large peak corresponding to $\omega_2$ of the silver sample holder. The relaxation rate $\sigma$ as a function of temperature extracted from TF-\mSR\ in various fields for \CS\ is plotted in Fig.~\ref{TF-muon}(c).

\begin{figure}[t]
\centering
\includegraphics[width=85mm]{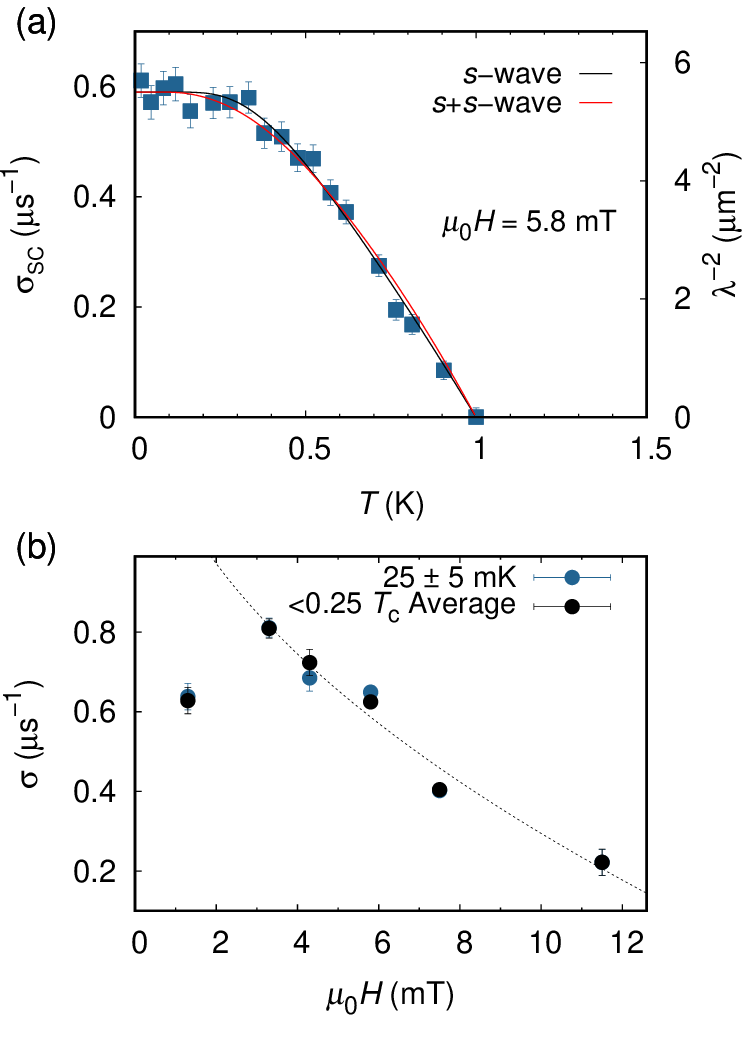}
\caption{(a) The temperature dependence of superconducting contribution to the relaxation rate $\sigma_{\rm{SC}}$ (green symbols), and the fits with a single $s$-wave gap and two $s$-wave gaps in black and red, respectively. (b) The muon Gaussian relaxation rate $\sigma$ as a function of applied magnetic field at base temperature ($\sim25 \pm 5$~mK) and the average at low temperature, below 0.25\Tc\ for each curve in Fig.~\ref{TF-muon}(c).}
\label{Penet}
\end{figure}

We extract \Tc\ from the TF-\mSR\ in various fields, where we define \Tc\ as the intersection of the line of best fit with sharpest slope in the transition seen in $\sigma$ and the normal state nuclear contribution $\sigma_{\mathrm{N}}\sim0.210$. We calculate the expected nuclear magnetic moment by first estimating the muon stopping site based on the Hartree potential, where we find the preferred muon stopping site is (0.65,0.75,0.15). Based on the magnetic active nuclei, only Sb in our case, we find an expected nuclear dipolar field of 3.5323~$\mu$N. This corresponds to a $\sigma_{\mathrm{N, calc}}\sim0.210~\mu$s, in agreement with the value measured experimentally, as shown in Fig.~\ref{TF-muon}(c). The applied magnetic fields, as extracted from the precession of muons in the Ag sample holder $\omega_{2}$, are plotted against \Tc\ in Fig.~\ref{Aniso}(c), and we fit the WHH relation to obtain the upper critical field from TF-\mSR\ as $10.5\pm0.4$~mT. This \mHct\ value is consistent with estimates based on 50\% drop in resistivity and those extracted from magnetization measurement with field applied perpendicular to the $ab$-plane.

From the field dependence of $\sigma$ measured well below \Tc\ at $25\pm5$~mK, shown in Fig.~\ref{Penet}(b), we find a peak at low fields. The FLL state is only realized at fields well above this peak region, where ideally in strong type-II superconductors we expect a relatively weak field dependence above this peak and below the upper critical field. The temperature dependence for TF-\mSR\ measured in 1.3 mT field, shown in Fig.~\ref{TF-muon}(c), shows a clear peak below \Tc\ consistent with the field being below \mHco . Since \CS\ is barely type-II, we do not have a wide range of weak field dependence, but nevertheless choose 5.8~mT, which is well above the peak position, as a field representing the highest likelihood of realizing a homogeneous FLL state.

For fields approaching the upper critical field, $\left[H / H_{c 2} = 0.5\right]$, the penetration depth can be calculated from the relaxation rate using the Brandt formula~\cite{brandt2003properties} for a triangular Abrikosov vortex lattice:

$$
\sigma_{\mathrm{sc}}(T)=\frac{0.0609 \times \gamma_{\mu} \phi_{0}}{\lambda^{2}(T)} 
$$
\noindent
where $\sigma_{\mathrm{sc}}(T)$ is in $\mu \mathrm{s}^{-1}$ and $\lambda(T)$ is in nm. $\phi_{0}$ is the magnetic flux quantum, $\left(2.067 \times 10^{-15} \mathrm{~Wb}\right)$.

We can relate the temperature dependence of the relaxation rate to the penetration depth with
$$
\frac{\sigma_{s c}(T)}{\sigma_{s c}(0)}=\frac{\lambda^{-2}(T)}{\lambda^{-2}(0)}.
$$

The temperature dependence of the energy gap $\Delta(T, \hat{k})$ within BCS theory~\cite{carrington2003magnetic} is given by:
$$
\Delta(T, \hat{k})=\Delta(0) \tanh \left\{1.82\left(1.018\left(\frac{T_{c}}{T}-1\right)\right)^{0.51}\right\} g_{\hat{k}}
$$
\noindent
where $\Delta(0)$ is the gap magnitude at zero temperature, and the $g_{\hat{k}}$ term accounts for the orientation $(\hat{k})$ dependence of the gap function, which can, for example, be substituted with 1 for an $s$-wave model and $|\cos (2 \phi)|$ for a $d$-wave model, where $\phi$ is the azimuthal angle.

\begin{figure}[h]
\centering
\includegraphics[width=85mm]{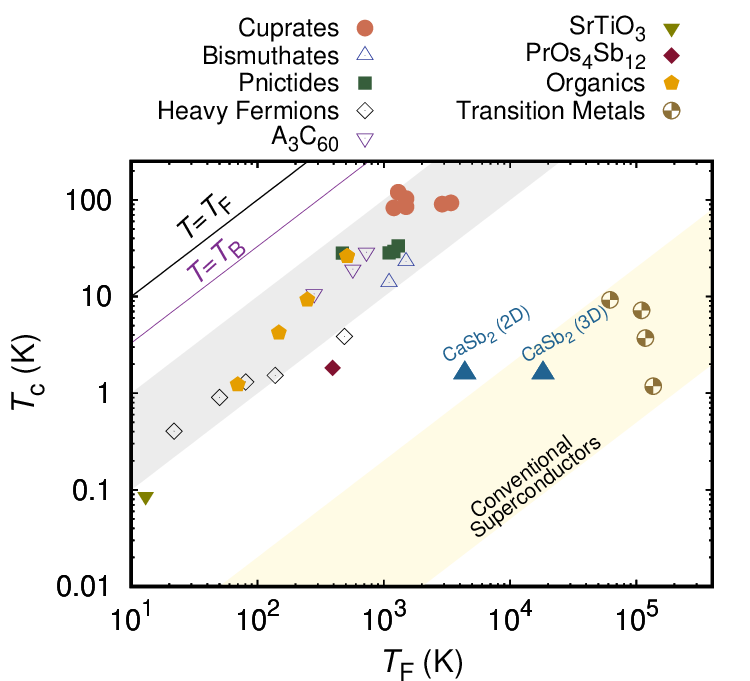}
\caption{Uemura plot showing the superconducting transition temperature \Tc\ vs the Fermi temperature $T_{\rm{F}}$, where \CS\ is shown as blue triangle assuming a 2D and 3D nature of charge carriers. The Bose-Einstein condensation temperature is shown as the $T_{\rm{B}}$ line and the Fermi temperature as the $T_{\rm{F}}$ line. Data for materials in literature are also plotted~\cite{uemura2004condensation,poole2007superconductivity,carbotte1990properties,peabody1972magnetic,adroja2018multigap,albedah2017absence,bhattacharyya2018brief,sonier2007hole,uemura1991muon,talantsev2015universal,uemura1989universal,lin2013fermi}. We highlight the region of unconventional superconductors in grey at the top left region and conventional superconductors in the bottom right region. }
\label{Uemura}
\end{figure}

\CS\ has a coherence length over normal state mean free path of about 1.78~\cite{oudah2022superconductivity}, which places it at the border between clean and dirty limit. The temperature dependence of the superconducting gap can be obtained from the temperature dependence of the penetration depth in the clean limit with the relation

$$
\frac{\lambda^{-2}(T)}{\lambda^{-2}(0)}=1+2\left\langle\int_{|\Delta(T, \hat{k})|}^{\infty}\left(\frac{\delta f}{\delta E}\right) \frac{E d E}{\sqrt{E^{2}-\Delta^{2}(T, \hat{k})}}\right\rangle
$$
\noindent
while in the dirty limit with the relation

$$
\frac{\lambda^{-2}(T, \hat{k})}{\lambda^{-2}(0)}=\left\langle\frac{\Delta(T, \hat{k})}{\Delta(0)} \tanh \left[\frac{\Delta(T, \hat{k})}{2 k_{B} T}\right]\right\rangle
$$
\noindent
where $f=\left[1+\exp \left(E / k_{B} T\right)\right]^{-1}$ is the Fermi function, and quantities in brackets are the average over the Fermi surface. 


Considering previous specific heat measurements showing deviation from single-gap BCS~\cite{oudah2022superconductivity} and that tunnel diode oscillator measurements on \CS\ are better fitted with a two-gap model, we utilized a two-gap model fit for our muon data where the total depolarization is expressed as the sum of two components:

$$
\frac{\sigma_{F L L}^{-2}(T)}{\sigma_{F L L}^{-2}(0)}=x \frac{\sigma_{F L L}^{-2}\left(T, \Delta_{0,1}\right)}{\sigma_{F L L}^{-2}\left(0, \Delta_{0,1}\right)}+(1-x) \frac{\sigma_{F L L}^{-2}\left(T, \Delta_{0,2}\right)}{\sigma_{F L L}^{-2}\left(0, \Delta_{0,2}\right)}
$$
\noindent
where $\Delta_{0,1}$ and $\Delta_{0,2}$ are the gap values at zero temperature and $x$ is the fraction of gap-1 over the sum of two gaps. We fit the gap using an $s+s$ wave model as shown in Fig.~\ref{Penet}(a). Assuming the zero-field tunnel diode oscillator measurement performed on \CS\ is representative of our samples, supported by the similarity of the specific heat data reported in the same paper~\cite{duan2022nodeless} with our specific heat measurement on our samples~\cite{oudah2022superconductivity}, we accept the presence of two gaps in zero field in \CS . Two gaps with the same value as reported $(\Delta_1(0)/k_B$\Tc = 1.8 and $\Delta_2(0)/k_B$\Tc = 0.81)~\cite{duan2022nodeless} and single gap $(\Delta(0)/k_B$\Tc = 1.59) fit are shown in Fig.~\ref{Penet}(a) in red and black, respectively. 
We note that for data measured in 5.8 mT a single $s$-wave gap is sufficient to fit the data, and the two-gap model does not significantly improve the fit. Also, the evolution of two gaps with applied magnetic fields has been demonstrated in measurements on NbSe$_2$~\cite{sonier2007muon}, and a similar evolution of the two gaps with applied magnetic field may appear in other multi-gap superconductors, such as \CS . The relaxation rate measured in different fields, Fig.~\ref{Penet}(b), does not follow single temperature linear temperature dependence for the fields measured, which further supports the two gap scenario for \CS . More TF-\mSR\ measurements to confirm the change in field dependence of \CS\ is needed, as has been demonstrated for other two-gap superconductors such as LaNiC$_2$~\cite{sundar2021two}. Nevertheless, the TRS breaking detected in ZF-\mSR\ supports two gaps in \CS\ with a complex order parameter, further discussed in Sec.~\ref{ZF}.


The strong magnetic field dependence of $\left(\sigma_{s c}\right)$ in \CS , shown in Fig.~\ref{Penet}(b), is associated with the low \mHct\ compared with the applied fields, and may be related to the faster suppression of smaller gap to the superfluid density. Such behavior has been discussed in other two-gap superconductors such as NbSe$_2$~\cite{sonier2000musr}, MgB$_2$~\cite{serventi2004effect}, and SmFeAsO$_{0.85}$~\cite{khasanov2008muon}. We notice a peak in the low-temperature field-dependence around 3.3~mT, where such a peak typically appears around \mHco .

\begin{figure*}[t]
\centering
\includegraphics[width=178mm]{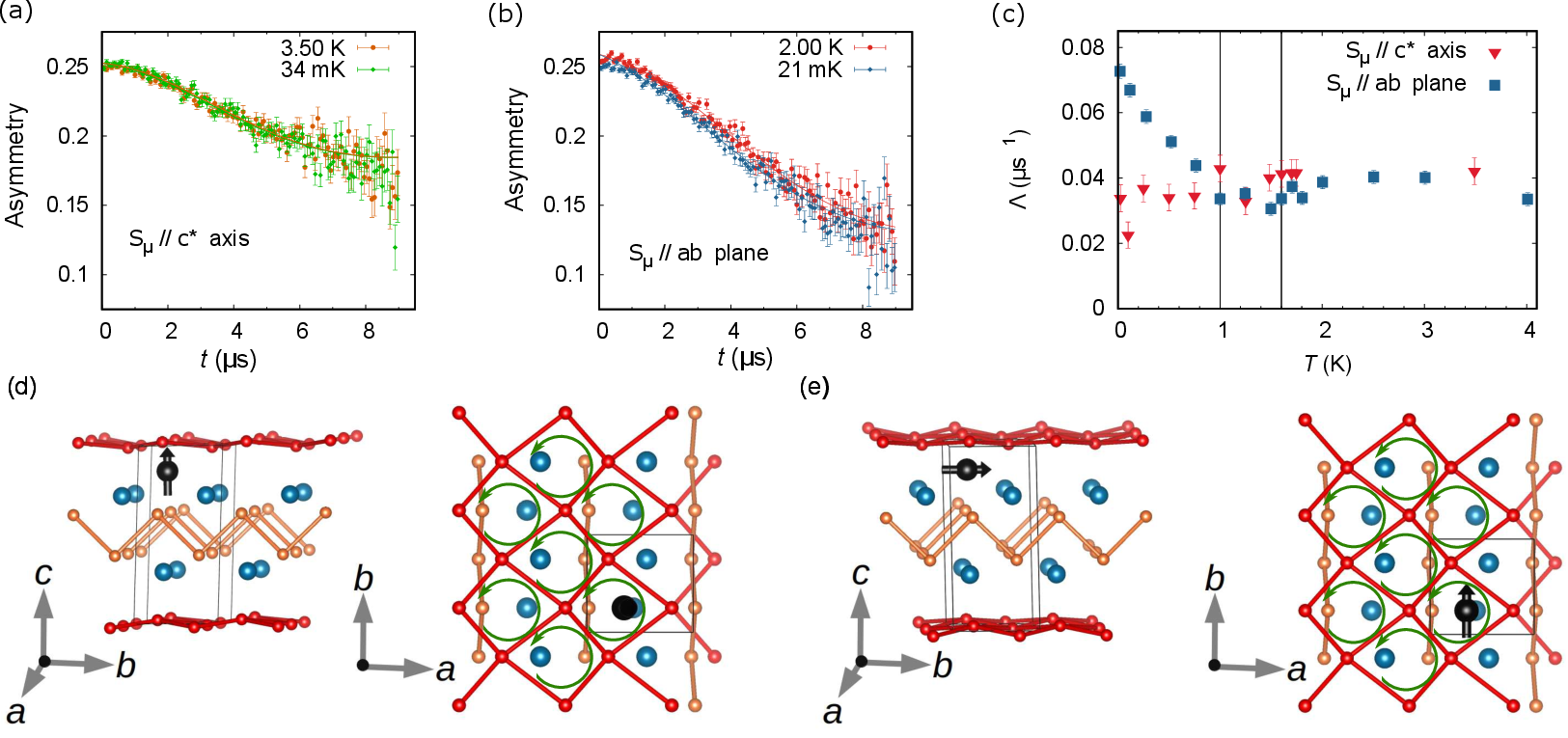}
\caption{(a) ZF \mSR\ spectra collected at 3.5~K and 34~mK for spin perpendicular to $ab$-plane (parallel to $c*$-axis), with fit using Kubo-Toyabe function. (b) ZF \mSR\ spectra collected at 2.00~K and 21~mK for spin parallel to $ab$-plane, fit using Kubo-Toyabe function. We can see a clear difference between the asymmetry shown in (a) and (b), indicating the detection of spontaneous fields developing in the superconducting state, but only seen by muons with spin aligned in the $ab$-plane. (c) Temperature dependence of the electronic relaxation rate $\Lambda$ with muon spins parallel to $ab$-plane and perpendicular to $ab$-plane (parallel to $c*$-axis). A clear increase in the extracted rate can be seen only for one orientation. (d) Muon stopping site inside the \CS\ crystal structure and spin direction for the experiment in (a) with $S_{\mu}//c*$-axis. (e) Same muon stopping site as (d), but with $S_{\mu}//ab$-plane as in experimental result of (b).}
\label{ZF-muon}
\end{figure*}

The likely presence of multiple gaps along with the possibility of gap anisotropy both can affect the temperature dependence of FLL, and the high field compared with upper critical field used in our TF-\mSR\ experiments make it difficult to makes a definitive statement on the gap symmetry based on our relaxation rate data. The relatively high field used in our TF-\mSR\ experiment makes corrections to extract the penetration depth from the relaxation rate data difficult, due to the likely distortions to the FLL state in our case. Nevertheless, we give an estimate using of the penetration depth $\lambda _{ab}= 426$~nm.

We compare the superconducting state in \CS\ with other known superconductors using the well-known Uemura plot. We plot the superconducting transition temperature against the Fermi temperature for \CS\ along with various other superconductors in Fig.~\ref{Uemura}, where we highlight the region of unconventional superconductors in grey at the top left region and conventional superconductors in the bottom right region. Considering the quasi-2D nature of \CS , we estimate the Fermi temperature assuming a 2D system via the relation $T_F=\frac{(\hbar^2\pi)n_{2D}}{k_Bm^*}$ and for a 3D system via the relation $T_F=(\hbar^2/2)(3\pi^2)^{2/3}n^{2/3}/k_Bm^*$~\cite{uemura2004condensation}. We use the previously reported carrier concentration and effective mass $m*$~\cite{oudah2022superconductivity}. Based on our estimates, \CS\ appears in a region in between conventional superconductors and unconventional superconductors, where the estimate assuming a 2D system falls closer to the unconventional region.
 
\begin{figure*}[t]
\centering
\includegraphics[width=178mm]{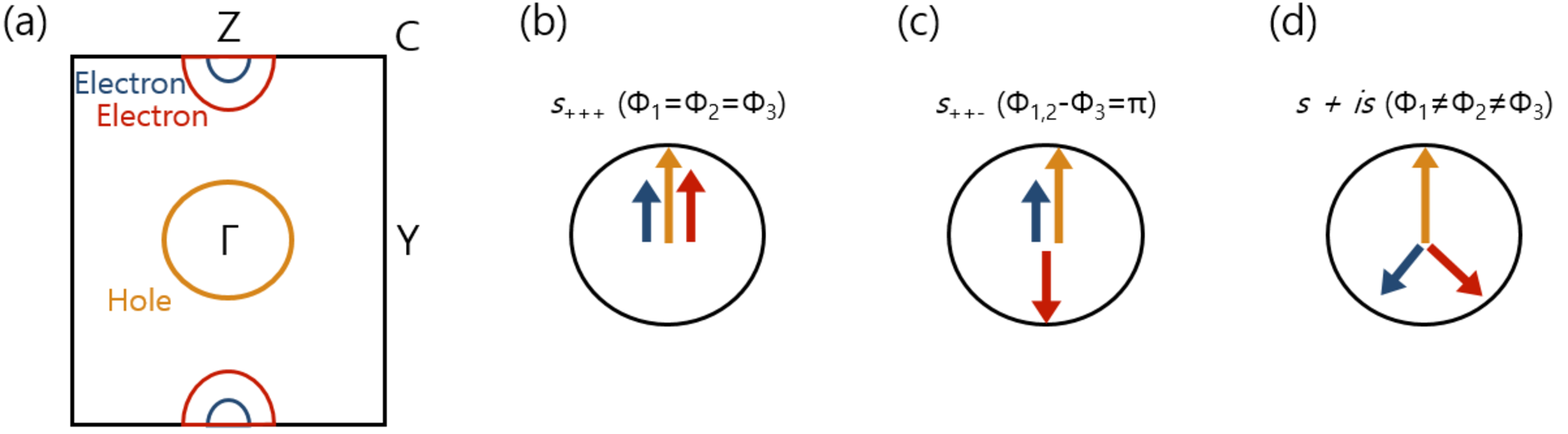}
\caption{\rev{Schematic of the Fermi surface topology of \CS\ shown in (a) with possible $s$-wave superconducting states shown in (b)-(d). A frustrated pairing $s+is$ state shown in (d) with an arbitrary phase shift between the components of the order parameter allows for breaking of TRS and in TF-\mSR\ the superconducting contribution to the relaxation rate $\sigma_{\rm{SC}}$ would follow a two-gap temperature dependence.}}
\label{GapSymmetry}
\end{figure*}

\subsection{\label{ZF} Zero-Field Muon and Time Reversal Symmetry Breaking}

\rev{Having characterized the superconducting state in the presence of a magnetic field, we turn to muon spin relaxation measurements in zero-field (ZF) to search for spontaneous magnetic fields associated with breaking of TRS in the superconducting state. ZF spectra for \CS\ collected above \Tc\ and at the lowest temperature $\sim30$~mK are shown in Fig.~\ref{ZF-muon}(a) and (b) for muon spins perpendicular to the $ab$-plane and parallel to the $ab$-plane, respectively. In the absence of any static electronic moments, the muon polarization decay is due to the randomly oriented nuclear magnetic moments, which are described by the Gaussian Kubo-Toyabe function $\mathrm{G}_{K T}(\mathrm{t})$}

$$
G_{\mathrm{KT}}(t)=\frac{1}{3}+\frac{2}{3}\left(1-\sigma^{2} t^{2}\right) \exp \left(-\frac{\sigma^{2} t^{2}}{2}\right)
$$
\noindent
where $\sigma$ reflects the width of the field experienced by muons due to nuclear dipoles.

We fit the ZF spectra with the following relaxation function

$$
A(t)=A_{1} G_{\mathrm{KT}}(t) \exp (-\Lambda t)+A_{\mathrm{BG}}
$$
\noindent
\rev{where $A_1$ is the sample asymmetry and $A_{B G}$ is the background asymmetry. The $\exp (-\Lambda t)$ term represents any additional relaxation in the sample, indicating broken TRS. The zero field muon spin relaxation rates of \CS\ with muon spin perpendicular and parallel to the $ab$-plane show the expected nuclear dipole contribution at temperatures above \Tc. However, there is a significantly higher $\Lambda$ at low temperature, but only when the muon spin is parallel to the $ab$-plane, shown in Fig.~\ref{ZF-muon}(b). Fig.~\ref{ZF-muon}(c) contrasts the temperature dependence of the relaxation $\Lambda$ in the two directions. While there is no discernible change in $\Lambda$ for muon spins aligned along the $c^{*}$-direction, there is a rise in relaxation for muon spins oriented in the $ab$-plane. Interestingly, this increase in $\Lambda$ shows a linear dependence at low temperature and seems to appear at 1.0~K, well below \Tc $\sim1.6$~K. }

\rev{We explore possible interpretations for this spontaneous TRS breaking, taking into account its dependence on the muon spin direction and the temperature being well below \Tc . Triplet pairing in the superconducting state is excluded by a Knight shift that drops below \Tc ~\cite{swavetakahashi}.   
Breaking of TRS has been reported below \Tc\ in another multigap superconductor, SrPtAs, where it was attributed to a grain boundary effect observable in polycrystalline samples with a multicomponent order parameter. This is unlikely in \CS\ due to the single crystal nature of the sample in our experiment.}


\rev{More complicated superconducting order parameters and phase diagrams appear in several systems. UPt$_3$ has multiple superconducting phases and also exhibits TRS-breaking near 1 K, as seen in \mSR~\cite{luke1993muon} and Kerr effect~\cite{schemm2014observation} measurements. 
A second superconducting phase emerging below \Tc\ has also been discussed for PrOs$_4$Sb$_{12}$, containing magnetic Pr$^{3+}$, where an enhancement of \mHco\ associated with the secondary phase is reported~\cite{cichorek2005pronounced}.
We considered the possibility of a secondary phase emerging at $\sim1$~K in \CS , so analyzed the \mHco\ estimates in Fig.~\ref{Hc1} by fitting the data above and below 1~K. We see a slight increase of estimated \mHco\ based on data below 1~K compared with that above 1~K when $H//c*$, while smaller difference appears for data above and below 1~K of \mHco\ with $H//ab$. 
The possible anomaly in \mHco\ in \CS\ is similar to that observed in PrOs$_4$Sb$_{12}$~\cite{cichorek2005pronounced}, although if present it is much weaker in \CS\ and is only observed with field applied $H//c*$. The appearance of TRS breaking with a spontaneous field in the $c*$ direction may be related to the change in the $T^2$ dependence in \mHco\ measured in the same direction. However, for both directions a single fit to \mHco\ over all temperatures falls within the error bars, so this is not yet clean evidence of another phase.}

\rev{ \CS\ lacks atoms with local moment and contains lighter elements than the examples noted above, so local magnetism and strong SOC are unlikely origins of the breaking of TRS. Instead, we suggest the topology of \CS\ in the normal state can explain the observation of TRS breaking. The topologically non-trivial Dirac nodal lines in \CS\ have been shown theoretically to support topological superconductivity with $B_g$ pairing symmetry, which has been termed nodal-line superconductivity~\cite{ono2021z}. Also, the topologically trivial $A_g$ symmetry is also supported in \CS ~\cite{ono2021z}, which is supported by the nodeless gap behaviour observed in our TF-\mSR\ measurements and previous works~\cite{duan2022nodeless,takahashi2021s}. This leads us to conclude that even if a second superconducting phase emerges in \CS\ at low temperatures it must have no extended nodes in its gap. Consider the three electronic bands in the normal state of \CS , shown schematically in Fig.~\ref{GapSymmetry}(a) depicting two electron pockets and one hole pocket, the possible pairing symmetries depend on the relative phase $\phi$ of the three order parameters. We have three possibilities: phase difference of 0 (Fig.~\ref{GapSymmetry}(b)), phase difference of $\pi$ (Fig.~\ref{GapSymmetry}(c)), and a phase difference other than 0 or $\pi$ (Fig.~\ref{GapSymmetry}(d).}

\rev{In the case of frustration of the order parameter due to competing interband interactions, a phase shift other than 0 and $\pi$ may be realized. The case shown in Fig.~\ref{GapSymmetry}(d) would result in a complex order parameter $s+is$ and would possess a fully gapped superconducting order parameter, consistent with our TF-\mSR\ results. The breaking of TRS is supported by this complex order parameter in \CS . If $s+is$ is realized in \CS , it is conceivable that domains of $s-is$ will equally likely form. At high temperatures fluctuations between these two domains may be too fast to be detected by muons, and at lower temperatures the fluctuations slow down and are detected in our \mSR\ experiment. This could explain the mismatch between \Tc\ and the observed onset of TRS-breaking. Furthermore, the constant increase down to the lowest temperature suggests these fluctuations are quantum in nature. Considering the muon stoppage site, shown in Fig.~\ref{ZF-muon}, we consider the possibility of loops that develop on the distorted square-net layer. Whether this fluctuation scenario is valid or a second superconducting phase emerges at a lower temperature distinct from \Tc\ remains to be clarified in future experiments. However, the absence of any clear anomaly in $C_{\rm{p}}$ at $\sim 1$~K leads us to conclude that the TRSB likely emerge at \Tc .}


\section{Conclusion}

\rev{In this Article, we presented magnetic susceptibility measurements of the upper and lower critical field anisotropy of \CS\, and studied the superconducting state using muon spin-relaxation. The temperature dependence of transverse-field relaxation can be fitted equally well with a single-gap model or two-gap model. A two-gap scenario is more likely considering the heat capacity~\cite{oudah2022superconductivity,duan2022nodeless} and TDO measurements~\cite{duan2022nodeless} on \CS\ are well fitted by a two-gap model. Zero-field relaxation shows little temperature dependence when the muon-spin is parallel to the $c*$-axis, but an increase in relaxation appears below 1 K when the muon-spin is parallel to the $ab$-plane. This evidence of time-reversal symmetry breaking is not accompanied by any other discernible anomaly around this temperature in other measurements.}

\rev{In various materials, the onset of TRS breaking coincides with \Tc\ under ambient conditions and its detection is independent of the direction of muon spins. Considering the anisotropy and two gap nature of the superconducting state and the emergence of direction-dependent TRS breaking detected at 1~K, we consider $s+is$ as the order parameter to reconcile TF-\mSR\ and ZF-\mSR\ results. We suspect currents being generated on the Sb distorted square-net layer considering the muon stoppage site and the spontaneous fields emerge at \Tc . Further investigation of \CS\ is necessary to clarify the origin of this TRS breaking, and any connection it may have to the band structure topology. Considering that the band structure of \CS\ is dominated by Sb atoms sitting at the two distinct sites in the material, investigation of other superconducting $M$Sb$_2$ antimonides with similar structures is of great interest.}

\subsection*{Single Crystal Growth and Characterization}
Single crystals of CaSb$_2$ were grown using an Sb self flux method as described in Ref.~[\citenum{oudah2022superconductivity}]. This yields shiny, plate-like crystals with dimensions up to $3\times3\times0.5$\,mm$^3$ that are stable in air for several weeks. Phase purity and orientation of the crystals were checked by X-ray diffraction (XRD) using a Bruker D8 with Cu K$\alpha_1$ radiation (1.54056~\AA ).

\subsection*{Magnetic Susceptibility Measurements}

DC magnetization measurements were done using a Magnetic Property Measurements System 3 (MPMS3) from Quantum Design, equipped with a $^3$He insert. Crystals were aligned to measure parallel to and perpendicular to the $ab$-plane of the crystals. A quartz sample holder with small amount of gee varnish was used for magnetization measured at temperatures above the superconducting state. A plastic straw was used for measurements with the $^3$He insert in the superconducting state.

\subsection*{Muon Spin Relaxation Measurements} 
Muon spectroscopy measurements were performed at the M15 beamline at TRIUMF's Centre for Molecular and Material Science, which is equipped with a dilution refrigerator. Multiple crystals were mounted on a silver plate with GE varnish, then the silver plate was attached to a cold finger utilizing copper filled grease for good thermal conductivity in the dilution refrigerator. We further secured the samples with a thin silver foil over the samples before mounting them into the dilution refrigerator. We achieved true-zero field for zero-field (ZF) \mSR\ using the method previously described by Morris and Heffner~\cite{morris2003method} to accurately detect any spontaneous fields. We used non-spin-rotated mode for the ZF-\mSR\ measurements and spin-rotated mode for the TF-\mSR\ measurements. ZF-\mSR\ was performed on two different samples mounted such that beam direction (spin-direction) is perpendicular and parallel to the $ab$-plane. In spin-rotated mode the muon spins are rotated perpendicular to the beam velocity, spin lying in the $ab$-plane, before landing in the sample, and the field is applied to the sample along the beam direction, perpendicular to the $ab$-plane. The \mSR\ data were analyzed with musrfit software~\cite{suter2012musrfit} to obtain physical parameters.

\section{Acknowledgements}
We thank TRIUMF staff for their technical support during muon experiment. We thank M.~Sigrist, M.~Franz, G.~Luke, Y.~Uemura, J.~Sonier, A.~Ramires, H.~Matsuura, M.~Ogata, and S.~Kivelson.  for discussion. 
MO acknowledges the support by Stewart Blusson Quantum Matter Institute and the Max Planck-UBC-UTokyo Center for Quantum Materials. JB, DAB, and MCA acknowledge the support by the Natural Sciences and Engineering Research Council of Canada (NSERC).

\bibliography{CaSb2}

\clearpage
\widetext
\begin{center}
\textbf{\large Supplemental Materials: \\Time-Reversal Symmetry Breaking Superconductivity in \CS\  }
\end{center}

\setcounter{equation}{0}
\setcounter{figure}{0}
\setcounter{table}{0}
\setcounter{page}{1}
\makeatletter
\renewcommand{\theequation}{S\arabic{equation}}
\renewcommand{\thefigure}{S\arabic{figure}}
\renewcommand{\thetable}{S\arabic{table}}
\renewcommand{\bibnumfmt}[1]{[S#1]}
\renewcommand{\citenumfont}[1]{S#1}

\begin{figure*}[h]
\centering
\includegraphics[width=158mm]{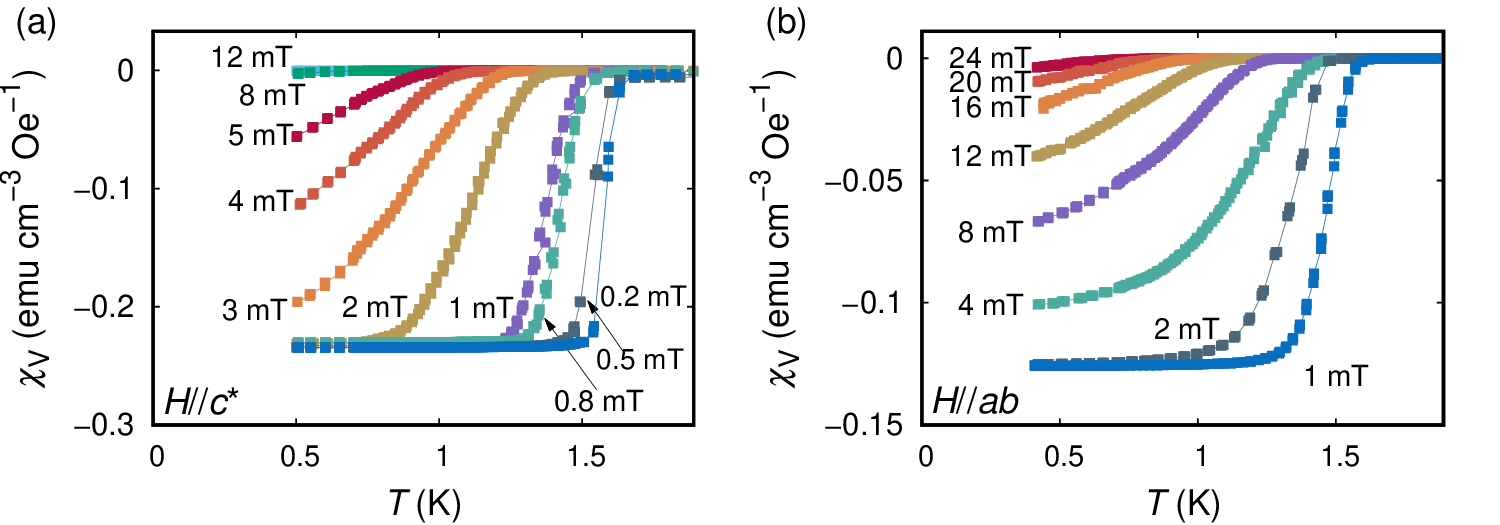}
\caption{Temperature dependence of the DC susceptibility measured with different applied fields using a zero-field-cooling procedure measured with $H//c*$ and $H//ab$ shown in (a) and (b), respectively. No demagnetization correction is applied.}
\label{UncorrectedVolumeSusc}
\end{figure*}

\begin{figure*}[h]
\centering
\includegraphics[width=108mm]{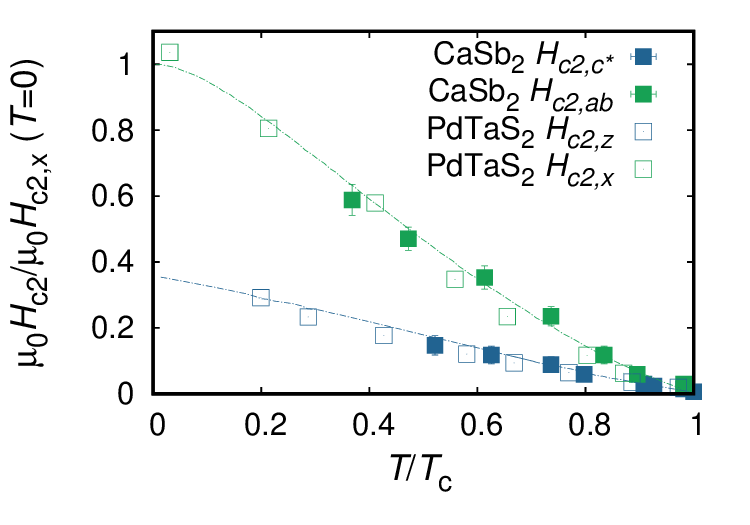}
\caption{Temperature dependence of the upper critical field for \CS\ normalized by upper critical field by $H//ab$ and \Tc . Temperature dependence of upper critical field for $s$-wave superconductor in a nodalline semimetal with asymmetric Dirac-crossing~\cite{endo2023theoretical}. Experimental data for the upper critical field
of nodalline semimetal PbTaSe$_2$~\cite{zhang2016superconducting} are shown with open symbols for comparison.}
\label{Hc2_nodal-line}
\end{figure*}

\begin{figure*}[h]
\centering
\includegraphics[width=178mm]{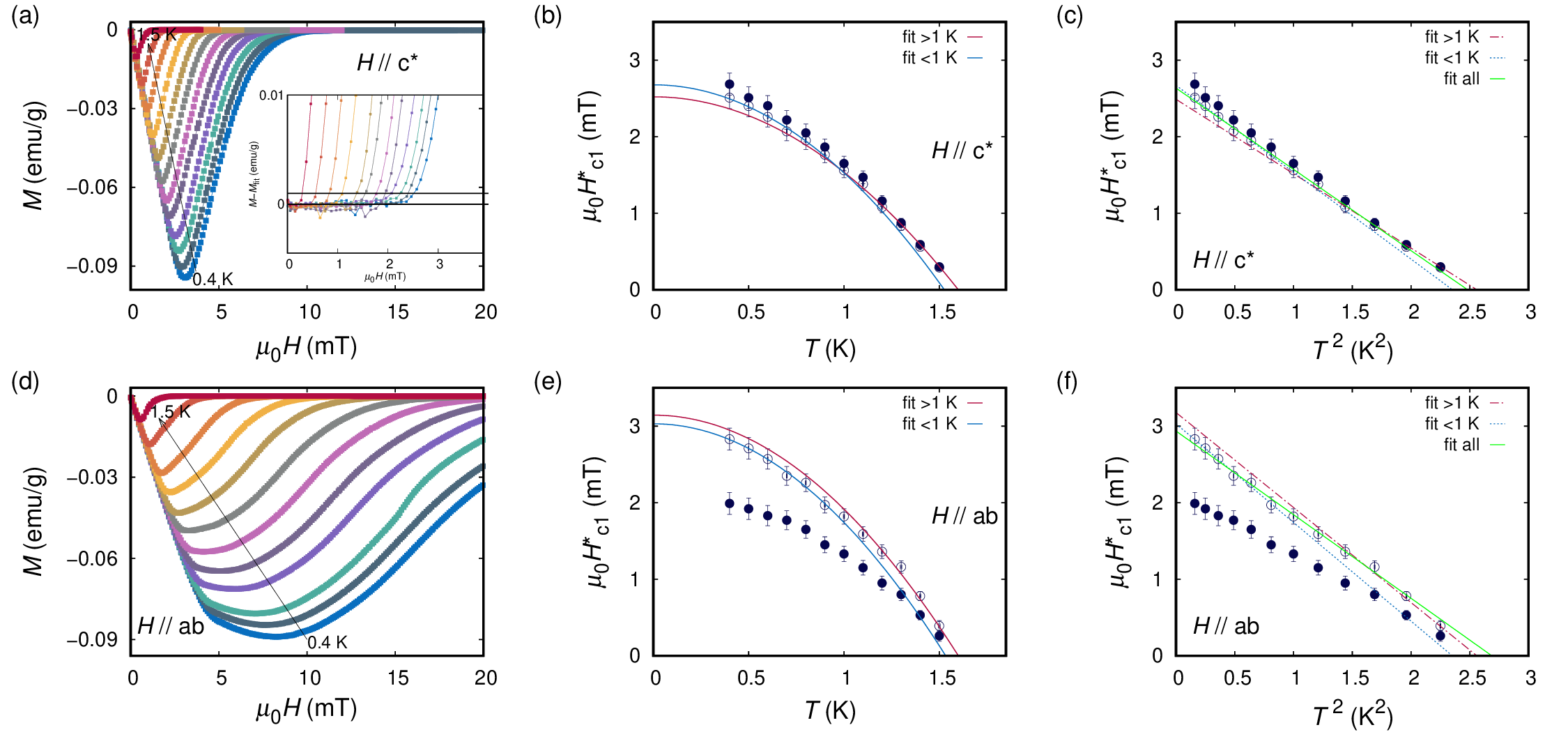}
\caption{The data shown in the Fig.~\ref{Hc1} of the main text are shown in (a) and (d). Measurements were performed on the same sample, but changing its orientation with respect to the applied field. Degaussing procedure was applied between measurements. The magnetization ($M$) as a function of applied field $\mu_0H$ measured at different temperatures $T$ below \Tc\ with field applied parallel to $c*$ direction and $ab$-plane shown in (a) and (d), respectively. Lower critical field \mHsco\ as a function of temperature ($T$) for $H//c*$ and $H//ab$, shown in (b) and (e), estimated using the magnetization data in (a) and (d) by setting 0.001~emu/g as significant deviation to decide \mHsco\ shown in open symbols. Method of linear fit used in the main text shown in filled symbols for comparison. (c) and (f) show \mHsco\ as a function of $T^2$ and fits to the data $<1$~K (dashed blue line), $>1$~K (dashed red line), and over the entire temperature range are shown (solid green line) for the open symbol data.}
\label{Hc1_Multi}
\end{figure*}

\begin{figure*}[h]
\centering
\includegraphics[width=178mm]{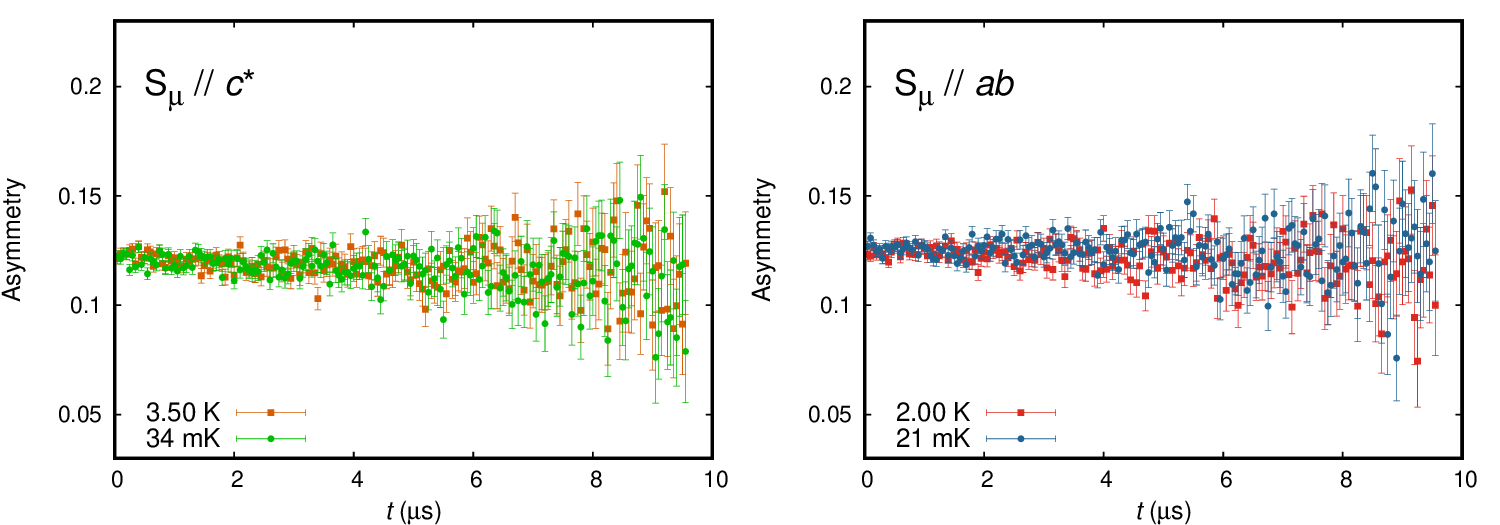}
\caption{The left-right counters for ZF \mSR\ spectra collected at 2.0~K and 21~mK for spin parallel to $ab$-plane (left) and $c*$-axis (right). Both spectra do not show any evidence of stray field.}
\label{LR-Counter-muSR}
\end{figure*}

\end{document}